\documentclass[manuscript]{rmaa}

\usepackage{paralist}

\usepackage{psfrag,color}
\usepackage[colorlinks=true,plainpages=true,citecolor=blue,linkcolor=blue, breaklinks=true]{hyperref}
\usepackage[none]{hyphenat}

\title{Kinematics of the Galactic Bubble RCW\,120} 

\author{
M. S\'anchez-Cruces, \altaffilmark{1}
A. Castellanos-Ram\'irez, \altaffilmark{2}
M. Rosado, \altaffilmark{3}
A. Rodr{\'i}guez-Gonz{\'a}lez, \altaffilmark{2}
and J. Reyes-Iturbide \altaffilmark{4}}

\altaffiltext{1}{Escuela Superior de F\'isica y Matem\'aticas, Instituto Polit\'ecnico Nacional}

\altaffiltext{2}{Instituto de Ciencias Nucleares, Universidad Nacional Aut\'onoma de M\'exico}

\altaffiltext{3}{Instituto de Astronom\'ia, Universidad Nacional Aut{\'o}noma de M\'exico}

\altaffiltext{4}{Divisi\'on de Mec\'anica. Tecnol\'ogico de Estudios Superiores de Tianguistenco, Estado de M\'exico, M\'exico}

\shortauthor{S\'anchez-Cruces}
\shorttitle{RevMexAA Main Journal Demo Document}

\listofauthors{M. S\'anchez-Cruces, A. Castellanos-Ram\'irez, M. Rosado, M., A. Rodr{\'i}guez-Gonz{\'a}lez \& J. Reyes-Iturbide}
\indexauthor{S\'anchez-Cruces}
\indexauthor{Castellanos-Ram\'irez}
\indexauthor{Rosado}
\indexauthor{Rodr{\'i}guez-Gonz{\'a}lez }
\indexauthor{Reyes-Iturbide}

\abstract{We studied the kinematics of the Galactic bubble RCW\,120 in the \mbox{[SII]$\lambda\lambda$6717,6731~$\rm \AA$} lines. We measured a LSR radial velocity ranging from $\sim$-74 to $\sim$-6~km~s$^{-1}$. We found evidence of expansion only in the northeast region of the nebula (from 20 to 30~km~s$^{-1}$). We found a high electron density around 4000 cm$^{-3}$ in the south-west region and we also found two arches-like structure indicating a density gradient.  We present 3D numerical simulations  of RCW\,120 using  {\sc Walicxe-3D} code in order to explore optical shell dynamics and its morphology. Our numerical results predict an average numerical electron density of the ambient  medium (in the southern region of the object) is between 3000 to 5000 cm$^{-3}$ in agreement with our values obtained from the observations. From our models, we do not expect X-ray emission coming from the external shell, due to the low expansion velocity value.}

\resumen{Presentamos el estudio cinem\'atico en las l\'ineas de \mbox{[SII]$\lambda\lambda$6717, 6731~$\rm \AA$} de la burbuja Gal\'actica RCW\,120. Determinamos que la velocidad radial de esta burbuja se encuentra entre $\sim$-74 to $\sim$-6~km~s$^{-1}$. Encontramos evidencia de expansi\'on  unicamente en la parte noreste de la nebulosa (de 20 a 30~km~s$^{-1}$). Encontramos alta densidad en la parte suroeste de la nebulosa (4000 cm$^{-3}$) as\'i como dos gradientes de densidad en la parte sur. Con la finalidad de explorar la din\'amica, morfolog\'ia y  emisi\'on \'optica de RCW\,120 presentamos simulaciones num\'ericas en 3D de esta burbuja usando el c\'odigo {\sc Walicxe-3D}. Los resultados num\'ericos predicen que en promedio la densidad del medio (en la parte sur del objeto) es de entre 3000 to 5000~cm$^{-3}$, lo que concuerda con los valores observados. Los modelos no predicen emisi\'on en rayos X provenientes del cascar\'on debido a la velocidad de expansi\'on baja.}

\addkeyword{H\,II regions}
\addkeyword{ISM: individual objects (RCW120)}
\begin{document}
\maketitle

\section{Introduction}
Massive stars ($\gtrsim$25 M$\odot$)  during their lives inject mechanical energy into the interstellar medium through their stellar winds (with a wind terminal velocity v$_{w}$=2000~km s$^{-1}$ and mass loss rate of $\dot{M} \approx$ 10$^{-6}$ to 10$^{-5}$~M$_{\odot}$ yr$^{-1}$; e.g. \citep{Friend-Abbott1986}). 
The stellar winds could sweep up the interstellar medium creating a dense shell filled with very hot gas known as wind bubble with a few 10 pc of size and superbubbles with sizes $\sim$100~pc in case of shell created by stellar winds of star associations. During the evolution of this bubble, the transfer of energy into the interstellar medium (ISM) can be carried out by radiative luminosity (L$_*$) and by mechanical luminosity due to winds (L$_w$=$\dot{M}v_w^2$).

The wind bubble formation around this stars begins when the massive star will first have formed an H\,II region in a very high density  with sizes of about 1 pc. During their main sequence and WR phases, the expansion of the ionized gas is driven by the hot shock stellar wind bubble.

In general the H\,II regions have spherical geometry in a homogeneous density medium, nevertheless some present asymmetric geometric when they are in inhomogeneous medium. This asymmetries can be classified according to their morphology: I) bow shocks of an ionizing star moving supersonically through a molecular cloud (Van Buren et al. 1990; MacLow et al. 1991). II) ``Champagne flow'' where it is assumed that the massive star is born in a medium with density gradients \citep{TenorioTagle1979}. III) bipolar morphology resulting  from the confinement of the ionized gas by a flattened structure of neutral gas and dust \citep{Campbell1984, Rodriguez1988}. 

A bubble created by a massive stars can be explained by the classical H\,II region.  One scenario indicates that the creation of bubbles is driven by the pressure difference between the ambient ISM and the ionized gas of the H\,II region. The analytical approximation to the expansion of an H\,II region in an homogeneous ambient medium and does not consider the stellar winds is given by \citep{Spitzer1978}. A second scenario is  the H\,II region expansion in presence of a stellar wind  in an uniform medium \citep{Dyson-Williams1980}. An example of the first case is presented by \citet{Martins2010}. They found that for the bubble RCW\,79, the mechanical wind luminosity is about $10^{-3}$ its ionizing luminosity in the RCW\,79 nebula indicating that the formation of the ring nebula could be attributed to the radiation pressure of the star. So that is in discussion the importance that the stellar winds of massive stars in the formation of the bubbles and superbubbles.
 
On the other hand, the standard model of \citet{Weaver1977} and \citet{Chu-MacLow1990} describes these bubbles as extended bubble structures of shock-heated gas emitting X-rays surrounded by a  shell of swept-up and cool ISM observed in optical emission. This model considered two shocks: The principal shock of swept-up ISM compressing in a thin shell to temperature about 10$^{4}$~K and a secondary shock of heated gas to temperatures 10$^{6}$-10$^{7}$~K. 

Considering this last scenery, it is to be expected that around massive stars observed a ring nebula. However, in the optical survey of Wolf-Rayet ring nebulae \citep{Heckathorn1982, Chu1983, Miller1993, Marston1994} shows that only $\frac{1}{4}$ from $\sim$150 observed Galactic Wolf-Rayet are associated with a ring-like nebulae \citep{Wrigge2005}. Also, in a recently work, \citet{Chu2008} and \citet{Chu2016} analyzed an HST observations of the H\,II region N11B (around an OB association LH10). The HST images of this H\,II region do not show visible ring-like morphological features, however the log slit spectrogram of [N~II]~$\lambda6583$ line shows splitting, indicating the existence of expanding shell of size of $\sim$15 pc and V$_{exp}$ of 15-20 km\,s$^{-1}$. The conclusion of \citep{Chu2016} is that the expansion velocity is associated with weak shocks within the photoionized medium. This medium does not produce an enough density jump to form a bright shell. If the medium is neutral, then there is a density contrast due to the strong shock, and is more easily to observe a ring nebula in HI emission than in optical.

Regarding the X-ray emission of superbubbles, is important take into account that the original work of \citet{Weaver1977} overestimated the observed X-ray luminosity in some bubbles. And for example, others studies like \citet{Harper2009} and \citet{Rogers2014} underestimate the X-ray emission (see \citep{Castellanos2015} for more information).  Therefore, from a theoretical point of view, the X-ray luminosity in a bubble is expected. And, seem to exist a correlation between the kinematic of the bubble and X-ray and optical emission. 

The study of the kinematics and  X-ray emission of the superbubbles presented by  \citet{Oey1996} it is proposed two superbubble categories in terms of dynamical data: high-velocity superbubbles and low-velocity superbubbles, the latter type being more consistent with the standard model, for high velocity is proposed an additional source of energy as a supernova explosion \citep{Rodriguez2011}. 
 
Also,  X-rays observations from superbubbles N70 and N185 located in the Large Magallanic Cloud (ideal laboratory to observe objects in X-rays by their low extinction), show X-rays emission inside the optical shell \citep{Reyes2014, Zhang2014} and present an excess of X-ray emission. Therefore, to explain the shell formation and the excess of X-ray emission \citet{Rodriguez2011} and \citet{Reyes2014} included an additional source of energy to explain them. 

On the other hand,  NGC 6888 is the only bubble in our Galaxy that has been observed in optical and X-ray emission \citep{Gruendl2000, Moore2000, Toala2016}, but the distribution of the X-ray emission could be described as filamentary, opposed to the center filled appearance expected in the analytic model.


In this work we address the problem of the correlation between kinematics and X-rays emission in the bubble RCW\,120 by using observations in the optical from the Fabry-Perot interferometer PUMA and X-ray data from CHANDRA, assisting us with hydrodynamic numerical simulations.

\subsection{RCW\,120}
RCW\,120 is a Galactic Bubble located at 1.3~kpc from the Sun \citep{Zavagno2007} bounded by a massive, dense shell with mass M$_{sh}\approx$~1200-2100~M$\odot$ 	\citep{Deharveng2009}. It is also called Sh 2-3 or Gum 58 and has a diameter of 1.9 pc \citep{Anderson2015}. The ionizing star of RCW\,120, CD-38$^{\circ}$11636 or LSS 3959 is an O8 type star \citep{Georgelin-Georgelin1970, Avedisova-Kondratenko1984, Russeil2003, Zavagno2007} located at $\alpha_{J200}$=17$^{h}$12$^{m}$20.6$^{s}$, $\delta_{J200}$=-38$^{\circ}$29'26''. Its magnitude in different filters is B=11.93 and V=10.79 \citep{Avedisova-Kondratenko1984}, with M$_{\star}\approx$30 M$_{\odot}$. K-band images of the ionizing star of RCW\,120 from 2MASS and SINFONI  suggest that the star is double being the companion of the same spectral type \citep{Martins2010}. Table~\ref{Table1} summarize the characteristics and properties of RCW\,120.

Current studies carried out with the NANTEN2, Mopra, and ASTE telescopes reveal two molecular clouds associated with RCW\,120 with a velocity separation of 20~km~s$^{-1}$ \citep{Torii2015}. The cloud at -28~km~s$^{-1}$ is distributed just outside the opening ring while the cloud at -8~km~s$^{-1}$ traces the infrared ring \citep{Anderson2015}. 
H$\alpha$~Fabry-Perot observations of RCW\,120 has been done with the survey CIGALE of the Milky Way and the Magellanic Clouds \citep{LeCoarer1993} on a 36-cm telescope (La Silla, ESO) by \citet{Zavagno2007}. These authors found a LSR radial velocity of ionized gas ranges -8 to -15~km~s$^{-1}$ and they did not find evidence for expansion of the ionized hydrogen associated with RCW\,120.

RCW\,120 is possible a triggered star formation region (i.e. like RCW 82, RCW 79) embedded in a molecular cloud \citep{Zavagno2007}.  2MASS, Spitzer and Herschel observations of this bubble reveal the existence of young stellar objects located in the massive condensation in the bubble surrounding \citep{Deharveng2009, Zavagno2010} indicating active star formation in the region. Since there should be shock compression triggering the star formation, we obtained Fabry-Perot data at [SII]$\lambda \lambda$6717,6731 $\rm \AA$ because it is easier to identify shocks in [SII] rather than in H$\alpha$, both became of the higher [SII]/H$\alpha$ line-ratio and also because [SII] velocity profiles of each velocity component are narrower than H$\alpha$ velocity profiles because of larger atomic weight of [SII] relative to the H$\alpha$.

The layout of the paper is as  follows. The observations and data reductions using the Fabry-Perot interferometer are presented in section 2. In section 3 we present the analysis of the kinematics of RCW\,120 as well as the morphology and density spatial distribution of this nebula into the ISM. In section 4 we show the numerical evolution of the X-ray emission and shell dynamics of bubble with the physical characteristics of RCW\,120. The discussion and conclusions are presented in section~5.

\section{Optical Observations and Data Reductions}

The observations were carried out in July 2014 using the 2.1~m  telescope of the Observatorio Astron\'omico Nacional of the Universidad Nacional Aut\'onoma de M\'exico (OAN, UNAM), at San Pedro M\'artir, B. C., M\'exico. We used the UNAM scanning Fabry-Perot interferometer PUMA \citep{Rosado1995}. We used a 2048$\times$2048 Marconi2~CCD detector with a binning factor of 4 resulting in a FoV of 10' and dimensions of \mbox{$512\times512$}\,pixels with a spatial sampling of 1''.3~pixel$^{-1}$. 

We obtained [SII]$\lambda \lambda$6717,6731 $\rm \AA$ Fabry-Perot data cubes of RCW\,120. The Fabry-Perot interferometer has a finesse of $\sim$24 leading to a sampling spectral resolution of 0.41 $\rm \AA$ (equivalent to a sampling velocity resolution of 19.0 km~s$^{-1}$ at 6717 $\rm \AA$) and  a free spectral range of 20~$\rm \AA$ (equivalent to a velocity range of 929~km~s$^{-1}$ at 6717~$\rm \AA$). The spectral resolution was achieved by scanning the interferometer free spectral range through 48 different channels producing velocity cubes of \mbox{$512\times512\times48$} \citep{Rosado1995}. The interference filter used was centered on 6721~$\rm \AA$) with a bandpass of 20 $\rm \AA$.  For calibrating the [SII] cube we used a Neon lamp (6717.04~$\rm \AA$ wavelength calibration) and the calibration data cube has dimensions of $512\times512\times48$. We  also obtained a set of direct images in H${\alpha}$ and [SII] using PUMA in its direct imaging mode (see Figure~\ref{Fig1}). The exposure time of each of the direct images was 120\,s. Observational and instrumental parameters are listed in Table \ref{Table2}.

The images were reduced using standard IRAF\footnote[1 ] {IRAF is distributed by 
National Optical Astronomy Observatory, operated by the Association of 
Universities for Research in Astronomic, Inc., under cooperative agreement with 
the National Science Foundation.}
routines. The data reduction and analysis of the
\mbox{Fabry-Perot} data cubes were performed using the CIGALE software. CIGALE  allows flat fielding correction,
wavelength calibration, construction of velocity maps and derivation of radial velocities, identification of sky-lines, profile extraction and fitting. In this case the data cubes
in the [SII] lines are not contaminated by line-sky emission. In the spectral window, the data cubes do not show any sky lines. No spatial or spectral smoothing was applied to the data.
The CIGALE data reduction process allows compute the parabolic phase map from the calibration cube. This map provides the reference wavelength for the line profile observed inside each pixel. Also, we can compute from the phase map, the wavelength, monochromatic, continuum maps. 

The extraction of the kinematic information from the Fabry-Perot data cube was done using the radial velocity map of RCW\,120. This map was obtained using the barycenter of the [SII]$\lambda$6717 $\rm \AA$ (using as the rest $\lambda$) velocity profile at each pixel. The radial velocity profiles were fitted by the minimum number of Gaussian functions after deconvolution by the instrumental function (an Airy function). The result of this convolution is visually matched to the observed profile. The computed width of the Gaussian functions is taken as the velocity dispersion (of the velocity component corrected by  instrumental function with a width of 38~km~s$^{-1}$).

\section{Data Analysis}
Figure~\ref{Fig1} shows the direct images of H${\alpha}$ and [SII]$\lambda \lambda$6717,6731 $\rm \AA$  emission-lines obtained with PUMA. The ionizing star is marked with a white arrow. We can see a well defined ionization front (IF) of RCW\,120 located in the southern region of the nebula showing a section of shell formed with dense material (shocked gas emitting in [SII] emission lines). This region corresponds to the dust condensations \#1 and \#7 \citep[see][]{Deharveng2009}  where more Young Stellar Object (YSO) are located. In the opposite side of the  IF it is observed an open structure.

\subsection{Kinematics of the Nebula}
Our Fabry-Perot data allow us to identify the global trends of the kinematics of RCW\,120 in [SII] emission. With the Fabry-Perot interferometer PUMA  it is possible to study both the large scale motions and  the punctual motions like expansion velocities unlike the classical long-slit spectrometers which is limited by the aperture spectrometry giving information of a small part of the object. 

Figure~\ref{Fig2} shows a mosaic obtained with the observed fields with the [SII] radial velocity profiles integrated over windows of 20$\times$20 pixels size (26''$\times$26'') overlaid in a [SII] image. The \textit{x} and \textit{y} axis indicates the (x,y) position of each integrated velocity profile. In this figure we can see that the  brightest emission is coming from the southern region of nebula, the two [SII]$\lambda \lambda$6717,6731~$\rm \AA$ lines are detected and are separated by about 15 channels (equivalent to 285~km~s$^{-1}$). Profiles from the northern region show different intensities between the two [SII] lines, whereas the profiles from the southern region seem to have the same intensity.  Since both lines $\lambda$6717~$\rm \AA$ and $\lambda$6731~$\rm \AA$ are emitted by the same gas, we will focus on the brighter line ([SII]$\lambda$6717~$\rm \AA$). 

We have fitted the velocity profiles shown in Figure~\ref{Fig2} using Gaussian functions and considering those with a significant signal to noise and taking into account the brighter line ([SII]$\lambda$6717~$\rm \AA$).  We corrected the observed FWHM to account for effects that broaden the lines. The instrumental broadening $\sigma_{inst}$ has been done in the deconvolution of the Airy function and in this case correspond to 38~km~s$^{-1}$. We computed the thermal broadening according to $\sigma_{th}$=$\sqrt{82.5(\rm T_4/A)}$ where T$_4$=T/10$^{4}$~K and A is the atomic weight of the atom. The correction  for thermal broadening to the [SII] is about 1.6~km~s$^{-1}$  (A$_{\rm sulfur}$=32.065u), compared with the instrumental broadening this has no effect upon our results. The fine structure of atoms broadening correction  $\sigma_{th}$ is important for hydrogen and helium recombinations lines, but not for metal lines such [SII] and [NII] \citep{Garcia2008}, so this correction is not required.  The turbulence broadening $\sigma_{tr}$ is related  other motions like nonthermal motions. Often complicated by superpositions along the line of sight of emission completely independent of the observed structure \citep{Courtes1989}, therefore the turbulence broadening has no effect upon our conclusions. 

From the velocity profile fit we found that the profiles in the northeastern region of the nebula (in the side of the open structure) are complex and broad, indicating the presence of different motions in that region. The profiles in the southern region of the bubble present single broad [SII] profiles corresponding to the radial velocity of the nebula. In the spectral window, the data cubes do not show any sky lines. No spatial or spectral smoothing was applied to the data.

In Figure \ref{Fig3} we present the spatial distribution of velocity profiles with single component (dark gray) and with double velocity components (light gray). As can be seen in this figure, the regions with two velocity component are located in the northeast side of the nebula (region of the open structure of the nebula) and some in the west side of the nebula. 

In order to show the differences between bright components we show in Figure \ref{Fig4} the radial velocity profiles of one region with a single velocity component and other one with two velocity components. The $x$-axis in the profiles is given in channels (each channel has associated a wavelength) and the $y$-axis gives the intensity of the line in arbitrary units. The single component of the velocity profiles correspond to the radial velocity of the nebula. In the case of velocity profiles with two components, these are called extreme velocity components. We shall refer to these as V$_{main}$ (represents the radial velocity of the nebula) and V$_{sec}$ (associated with others movements). These are marked with No.~0 and No.~1, respectively. 

Figure \ref{Fig5} display the velocity maps of the components obtained from
the fit profiles. Top panel shows the velocity map of the $V_{main}$ component. Bottom panel shows the velocity map of the V$_{sec}$  component. The highest LSR velocity values coming from  the southeast side of the nebula. The VLSR present a gradient of about $\Delta_{VLSR}$=60~km~s$^{-1}$ (varies between -74~km~s$^{-1}$ to -6~km~s$^{-1}$). This velocity gradient match with the CO distribution of the red cloud towards this nebula   presented by \citet{Torii2015}.


The expansion velocity determined for the regions with composite profiles is V$_{exp}$=20~km~s$^{-1}$, this value was calculated by using the expression: \mbox{V$_{exp}$=($V_{main}$-V$_{sec}$)/2}.

\subsection{Emission-line Ratio Maps and Electron Density}
Figure \ref{Fig6} shows the [SII]/H$\alpha$ line-ratio map. We can see that the highest values are located in the IF and from the northern region of the nebula showing the gas shocked on this regions. In the west region of the nebula the [SII]/H$\alpha$ values span from 0.2 to 0.3, while on the east region of the nebula, the values span form 0.1 to 0.2. These differences indicate that the ISM is not homogeneous. From Figure \ref{Fig6} we can see that the highest  [SII]/H$\alpha$ line-ratio values correspond to the northern region of the nebula where the profiles with highest velocity values are located.

From our [SII] data cube, where both 6717 $\rm \AA$ and 6731 $\rm \AA$ lines are detected, we were able to compute [SII]$\lambda$6717/[SII]$\lambda$6731 line-ratios (see Figure \ref{fig:nelec}). The line-ratio values span from 0.3 to 1.5. The lowest [SII]$\lambda$6717/[SII]$\lambda$6731 line-ratio values are located inside on the nebula.

In order to explore the morphology of the nebula RCW\,120, we obtained  the electron density map assuming a constant electron temperature of  10000~K, corresponding to the expected conditions in H\,II regions, using the [SII]$\lambda$6717/[SII]$\lambda$6731 line-ratio map.

The electron density map was computed using,
\begin{equation}
\label{eq1}
\frac{{\rm I}({\rm [SII]}\lambda6717)}{{\rm I}({\rm [SII]}\lambda6731)} = 1.49 \frac{1+3.77x}{1+12.8x}
\end{equation}
\noindent where $x$ is defined as $x \equiv$10$^{-4}$n$_e$T$^{1/2}$ and T is the electron temperature in units of 10$^{4}$ K~\citep{McCall1985}. 

In both cases we do not correct for differential extinction between H$\alpha$ and [SII] and the two lines of [SII]$\lambda\lambda$6717,6731 due to this lines are to close each other making the correction not significant.

In the lower panel of Figure~\ref{fig:nelec} we present the electron densities map. We obtained a maximum density of $\sim$3000~cm$^{-3}$ in the southern region of RCW\,120, and we fixed a minimum density of 0.03~cm$^{-3}$ (in order to run out numerical models). Also, from this figure we can see the spatial variation of electron density, n$_e$,  in the southern region of the nebula presents high electronic density ($\sim$3000~cm$^{-3}$) and it is possible to appreciate two arches-like structures (south-south-west and south-south-east). The external one present electronic density values between 3 to 400~cm$^{-3}$; while the internal arc has values between 3 to 40~cm$^{-3}$. The presence of two arches-like regions could be due to the difference in density of the clouds where the H\,II region seems to be evolving as proposed by \citet{Torii2015}.  The lower electronic density values are found in the northern region of the nebula; these differences in density cause a faster expanding material in this region allowing that the ionized gas could break the shell. This is known as the ``champagne phase'' \citep{TenorioTagle1979}.

\section{Chandra X-ray Observations and Gas Dynamic Simulation of RCW\,120}
Given the characteristics of the ionizing star of RCW\,120, it is expected that this bubble presents X-ray emission. In order to study their emission at these wavelengths, we obtained the X-ray data of RCW\,120 from the Chandra data archive. It has been observed with the Chandra X-Ray Observatory on 2012 june 30 and july 1 Chandra ObsID 13621 using the Advanced CCD Imaging Spectrometer (ASIS) with a exposure time of 49 $\rm ks$. We reprocessed level 2 data with {\it chandra-repro} and filtering out periods of high background using \textit{lc-clean}, left a total exposure of 39.7 $\rm ks$. We used Chandra Interactive Analysis of Observations (CIAO) version $4.7$ to analyze the X-ray data using calibration data from CALDB version~$4.6.8$. A  preliminary inspection  of  the  images obtained does  not  reveal  significant diffuse  X-ray  emission  within  the  nebula. Several point sources have been identified, even though there  is  no  detection  of  diffuse  X-ray  emission  coming from this bubble. 

In order to explore the X-ray emission and the shell dynamics of the RCW\,120, we developed 3D numerical gas dynamics simulations. In this case we do not  include photoionization nor radiation pressure because we have considered that even though massive stars form around of an H\,II region before the effect of its powerful wind interact with the surrounding medium (ionized gas), it is not entirely clear when one effect dominates over the other. As we mention above, in some cases, the creation of bubbles is simply driven by the pressure difference between the ambient ISM and the ionized gas of the H\,II region, and the effect of mechanical energy due to stellar wind is negligible.

Also, \citet{Mackey2016} compared synthetic infrared intensity maps made by numerical simulations with some Herschel infrared observations for RCW\,120. They conclude that, in order to obtain the shape and size of an brightly arc in the infrared waveband, they have to include stellar winds in their simulations.

On the other hand, \citet{Martins2010} point out that the dust emission in 24 $\mu$m (the same wavelength used in \citealt{Mackey2016}) does not completely favor a strong influence of stellar winds.

Recently, \citet{Gvaramadze2017} reported both, observations and numerical simulations of an H\,II region IRAS 18153-1651. They found an optical arc near to the centre of the nebula, and suggest that this arc is the edge of a wind bubble together with the H\,II region produced by a B star. They validated their hypothesis with analytical calculations of both, the radius of the bubble and that of the H\,II region which fit with the observations. Also, they obtained synthetic H$\alpha$ and 24 $\mu$m dust from 2D numerical simulations of radiation hydrodynamics. Their results match in good agreement  with the observations, in the case of the morphology and surface brightness.

Therefore, in order to support the use of only stellar wind in
our simulations, we follow the work of \citet{Raga2012}. They obtained an analytical model for an expanding H\,II region driven by  stellar winds and the ionizing radiations, both of them coming from the central source. The transition between the two phases is related with the $\lambda$ parameter (see equation (29) in \citealt{Raga2012}. and equation (26) in \citealt{Tinoco2015}):

\begin{equation}
\lambda=63\left(\frac{\dot{M}}{5\times 10^{-7}\,\textrm{M}_{\odot}\,\textrm{yr}^{-1}}\right)\left(\frac{v_{w}}{2500\,\textrm{km\,s}^{-1}}\right)^{2}\left(\frac{n_{a}}{10^{5}\,\textrm{cm}^{-3}}\right)^{1/3}\left(\frac{10^{49}\,\textrm{s}^{-1}}{S_{*}}\right)^{2/3}\left(\frac{1\, \textrm{km\,s}^{-1}}{c_{0}}\right)^{1/3}
\end{equation}

Basically, they conclude that if $\lambda > 1$, then the expansion of the region becomes the model for a wind-driven shell with a negligibly thin H\,II region.

In such way, we have calculated the $\lambda$-parameter in terms of the values used in our numerical simulations (see Table~\ref{Table3}), we obtained a $\lambda >1$. Therefore, in a good approximation, we can only consider the effects of stellar wind in order to study the X-ray emission in our numerical models.

Finally, we perform a set of three different numerical simulations using  the {\sc Walicxe-3D} code \citep[see][]{Esquivel2010, Toledo-Roy2014}. This code solves the hydrodynamic equations on a three dimensional Cartesian adaptive mesh using a second-order finite volume conservative  Godunov upwind method, with HLLC fluxes \citet{Toro1994} and a piecewise linear reconstruction of the variables at the cell interfaces with a Van Leer slope limiter. Additionally, the code includes artificial viscosity for the purpose of stabilize the simulation. The energy equation includes the cooling function appropriate for describing the cooling of the shocked wind material. The cooling function, for different metallicities, was obtained from the freely available  {\sc chianti} database \citep{Dere1997, Landi2006}.

The computational domain is a cube of 5 pc on a side, with an uniform medium of number density $n_0$ and temperature $T_0$ (see below).  We used 5-levels in the adaptive grid with a maximum resolution of 256 points along each of the x, y and z axes. 

Our numerical models considered appropriate values of the stellar wind 
velocity and mass injection rate for one single O8V star (see \citealt{Sternberg2003} \& \citealt{Mackey2015} for the mass used in the simulation). The star was placed at the center of the box simulation and its stellar wind was imposed in a region of 5 pixels, corresponding to a physical radius of 0.08 pc. {In this region we impose a steady-state spherical stellar wind solution with a $\propto R_w^2$ density profile, such tat: $\dot{M}=4\pi\rho V_{\infty}R_w^{-2}$. Also, the energy injected by the stellar wind must satisfies $E=\rho \vert \mathbf{u}\vert^2/2+ P/(\gamma-1)$, with $P$ the pressure due to the wind. Moreover, in order to improve the steady-state wind, we include a slope in velocity such that $v_{\infty}\propto R_w^{-1}$.

We explored three values of $n_0$, the cloud density, these are: 2000, 3000 and 6000~cm$^{-3}$. Table~\ref{Table3} shows the initial condition physical properties for the  three numerical models, M1, M2 and M3. Column~1 indicates the model. Column 2 shows the wind velocity. Column 3 shows the mass loss rate. Columns 4 and 5 show the  is the density and temperature for the surrounding environment. Column 6 shows the luminosity. We used a super-solar metallicity for the wind and a sub-solar metallicity in the interstellar medium, 3 and 0.3~Z$\odot$, respectively \citep[see][]{Castellanos2015}. We have carried out time integrations from $t = 0$  up to $t = 0.4$~Myr for all the models.

\vspace{2cm}
 \subsection{Numerical Results}

The evolutionary X-ray emission for the models was  computed using emission coefficients in the low-density regime taken from {\sc chianti} data base as function of the metallicity (see Figure 2 in \citealt{Castellanos2015}) in the band of 0.6 to 10~keV (soft and hard \mbox{X-ray} band).
Figure~\ref{fig:nummaps} shows the temperature, column densities and
intrinsic X-ray emission maps, upper middle and lower panels, respectively, for the models M1, M2 and M3, left, center and 
right panels, respectively, at t=400~kyr. From row density maps, one can see the radius of the shell driven by the stellar wind. As we expect the model M3, a denser interstellar medium, have small shell radius. On the other hand, we can not see a contribution of the X-ray emission coming from the shocked environment, by the leading shock. This shell, in the case of RCW\,120  is highly dense and the expansion velocity is around 20-40~km~s$^{-1}$. The gas, in the post-shock region, has a temperature around tens of thousands kelvins, i.e. with optical emission. Nevertheless, a weak X-ray emission region is predicted in the not shocked wind region, the very low dense gas region (see also the upper panel in Figure~\ref{fig:numgraphs}).

In addition, the time when the models reach their maximum X-ray luminosity value is different from every model, as one would expect. It is about 60~kyr for model M1, 90~kyr for model M2 and 210~kyr for M3. Nevertheless, the intrinsic X-ray emission is about two order of magnitude less than the X-ray emission coming from the central star associated to RCW\,120 (as we can see in the lower panels of Figure~\ref{fig:nummaps}). In this way, the X-ray luminosity in this region has a very low value to be detected.

Therefore, from our numerical simulation we do not expect a detectable X-ray emission coming from the shocked gas of RCW\,120  even when the temperature values reach $10^{7}$ K.
Table \ref{Table5} shows the results for for models M1, M2 and M3. Column 1 shows indicates the model. Columns 2 and 3 show the position and width  of the optical shell (the shocked interstellar medium). Columns 4 shows the total soft X-ray luminosities  at t=400~kyr and Column 5 shows the maximum value of the soft X-rays luminosity reached by this numerical bubbles  in the band: 0.6 to 10 keV.

Notice that the external shell, in our models, do not have X-ray emission despite that the main source of X-ray emission in bubbles is the 
medium swept up by the front shock.  The shock wave rises the temperature of the gas and the maximum temperature, the post-shock 
temperature, is around $10^7$~K for stellar wind velocities of $\sim$10$^3$~km~s$^{-1}$. In order to see this behavior more clearly, we present the density, temperature and velocity behavior as a function of the radius in Figure~\ref{fig:numgraphs}. We notice, as we mention previously, that the zone of very low density corresponds to high temperature as well as high velocity. When the density increases (upper panel in Figure~\ref{fig:numgraphs} ), the other variables go down (middle and lower panels in Figure~\ref{fig:numgraphs}). On the other hand, as we can see from Figure~\ref{fig:numgraphs} (bottom panel), the numerical value of the velocity obtained from the simulations, $\sim$4 km s$^{-1}$ for  the RCW 120 is in good agreement with the expansion velocity measured from the observational data, which is in the range 20-40~km~s$^{-1}$, at a radius of 1.9 pc.

On the other hand, the density of the external shell is, 
at least, 4 times larger to the interstellar medium density, for an adiabatic expansion of a supersonic gas. However, the case of bubble evolving 
into dense ambients, i.e. the  maternal cloud, the external shell of this bubble lose an important fraction of their thermal energy because of 
the radiative processes (cooling). The temperature of the gas, into the outer shell, drops from $10^7$~K to $10^4$~K in a cooling 
time giving by,

\begin{equation}
\label{eq:tcool}
t_{cool}=\frac{E_{th}}{L_{rad}}=\frac{n k T/(\gamma-1)}{n^2 \Lambda(Z,T)}=\frac{k T/(\gamma-1)}{n \Lambda(Z,T)},
\end{equation}

where, $E_{th}$ is the thermal energy, $L_{rad}$ is the cooling rate, $k$ is the Boltzman constant, $n$ is the numerical density, $T$ the 
gas temperature, $Z$ is the metallicity of the gas, and $\Lambda(Z,T)$ is the cooling function as a function of Z and T. We can also re-write the 
cooling time (for a gas with initial temperature of $10^7$~K) as,
\begin{equation}
\label{eq:tcoolbien}
t_{cool}=2.5\times 10^{4}\left(\frac{n}{1000}\right)^{-1} [yr]
\end{equation}

where, the cooling function values were obtained from \citet{Castellanos2015}. Using the equation \ref{eq:tcoolbien} 
we calculated the cooling time $t=1.25\times 10^{4}$, $8.34\times 10^{3}$, and $4.17\times 10^{3}$~yr, for the models M1, M2 and M3, respectively. Therefore, in the case of RCW\,120, we 
expect that the gas within  the external shell keeps at a temperature around tens thousands of kelvins.

Therefore, in order to reproduce the shell size and width, our simulation predicted the average numerical density of the ambient 
medium (in southern region of the object) between 3000-5000 cm$^{-3}$ and an interstellar medium density around 1000 cm$^{-3}$ for the northern region of RCW\,120.

\hspace{3cm}

\section{Discussion and Conclusions}
We have analyzed the kinematics and the predicted X-ray luminosity of the  galactic Bubble RCW\,120 using the PUMA Fabry-Perot interferometer and 3D numerical simulations. From the PUMA direct images we found that the
H$\alpha$ and [SII] emission (see Figure \ref{Fig1}) show in both cases a diffuse ``halo'', but in the [SII] image shows the ionization front in the southern region of the nebula. In both cases the halo is not perfectly symmetric and is more extended toward the northern region of the nebula. We found that the higher [SII]/H$\alpha$ ratios are located in the northern region of the nebula. 

Kinematic information of RCW\,120 in [SII] line emission obtained here reveals that the LSR radial velocity ranges from \mbox{$\sim$-74} to $\sim$-6~km~s$^{-1}$, in agreement with the values derived from H$\alpha$ by \citet{Zavagno2007}. Also, indicates that it may be a champagne phase. Double component profiles are present in the 
northern region of this nebula. On the southern region of the bubble no expansion velocity has no component; while in the northern region the expansion velocity span 20 to 30~km~s$^{-1}$. This behavior can be related with a complex structure in the interstellar medium where the shell is evolved, i.e., part of this bubble is within its molecular cloud.

We presented a density map obtained from the [SII]$\lambda6717$/[SII]$\lambda6731$  line-ratio emission, and we found a maximum density around 4000 cm$^{-3}$ in few regions of the southern region of RCW\,120 region. We also proposed the existence of two arches-like structure in southern region of the nebula emitting in [SII] emission lines. With densities between 3 to 400~cm$^{-3}$ (for the external arc) and between 3 to 40~cm$^{-3}$ (for the internal arc).  This is in agreement with the fact that there are  two molecular clouds physically associated with RCW\,120 with a velocity separation of 20~km~s$^{-1}$ \citep{Torii2015}.

Regarding to X-ray simulations, for all models we considered super-solar metallicity for the wind and a sub-solar metallicity in the interstellar medium, 3 and 0.3 Z$\odot$, respectively. The mass loss rate is $\dot{M}$=2.7$\times$10$^{-7}~$M$_{\odot}$~yr$^{-1}$ evolved at age of 0.4 Myr. We found that Model M2, with a numerical density in a single cloud of 3000 cm$^{-3}$ is the model that best fits the dynamics of the southern region of the bubble. This model predicts a R$_{Shell}$=2.12 pc, $\bigtriangleup$R$_{Shell}$=0.68 pc and no X-ray emission is predicted.
 
Moreover, we have considered an homogeneous density for the three models overlooking the champagne flow show in the case of RCW\,120. As far as we know, both in terms of models of H\,II regions, or of winds of stars, the case of RCW\,120 has not been simulated using a champagne flow. We expect this fact certainly reduce the pressure in the bubble (by providing a channel for it to escape), what will produce changes in the density and temperature. Nevertheless, following the morphology of RCW\,120, which is almost spherical, we expect this changes are not so relevant and we can speculate that the champagne flow effect is not determinant at all yet for the X-ray emission, at least in the southern region of RWC\,120, which is the densest region with the highest temperature, that most contributes to the aforementioned emission.

Finally, according to \citet{Crowther2007}, the minimum mass that a star has to be to become a WR star is $\sim$~25~M$_{\odot}$ (at solar metallicity), that means the stellar mass of the ionizing star in the case of RCW\,120 is enough to become a WR star. Therefore, we would expected that in the case of a more massive star, for example a Wolf-Rayet star or a Of star with high stellar winds, we would expected that the expansion velocity of the bubble obtained from the simulations presented in this work, will be higher than that found in RCW\,120 with a O8V star. Nevertheless,  it is important that taking into account that this effect depends of the medium as discuss in \citet{Chu2008} and \citet{Chu2016}.


\hspace{3cm}

\acknowledgments{We acknowledge support from CONACyT grants 253085 and 167625. This work was also supported by DGAPA-UNAM grants  PAPIIT-IN103116, IA-103115, IN-109715 and IG-RG-100516.  

Mónica S.C. acknowledges CONACYT for a doctoral scholarship. 

Based upon observations carried out at the Observatorio Astron\'omico Nacional on the Sierra San Pedro M\'artir (OAN-SPM), Baja California, M\'exico.}

\newpage

\begin{figure*}
\includegraphics[width=\columnwidth]{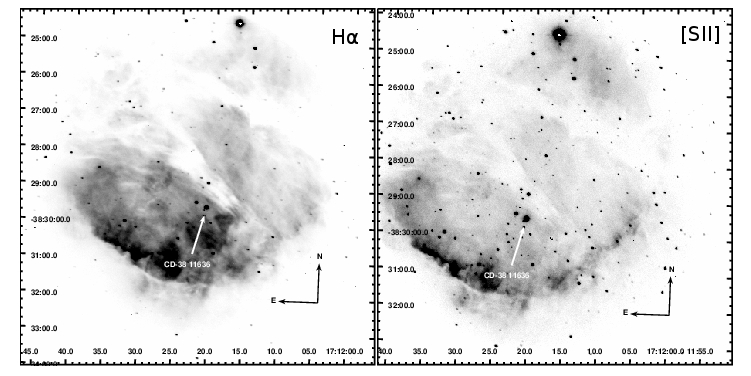} 
\caption{Direct images of RCW\,120, {\it left panel:}  shows the H$\alpha$ emission and {\it right panel:} shows [SII]$\lambda \lambda$6717,6731 $\rm \AA$  emission. We are pointing, with a white arrow, the ionizing star position (Sh2-3).}
\label{Fig1} 
 \end{figure*}

\begin{figure*}
\includegraphics[width=\columnwidth]{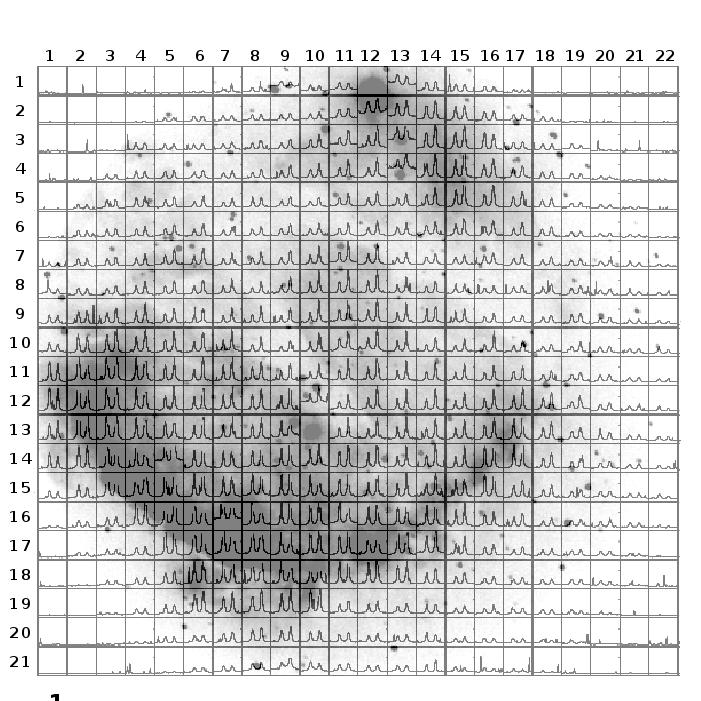} 
\caption{Some examples of the [SII] radial velocity profiles (20$\times$20 pixels velocity profile plots) of the different regions of RCW\,120 superimposed on a [SII] emission velocity map at -21~km~s$^{-1}$ (gray color). The numbers in top and left indicate the coordinates of the profiles position.}
\label{Fig2} 
 \end{figure*}
 
\begin{figure}[htp]\centering
\begin{center}
\includegraphics[width=1.0\columnwidth]{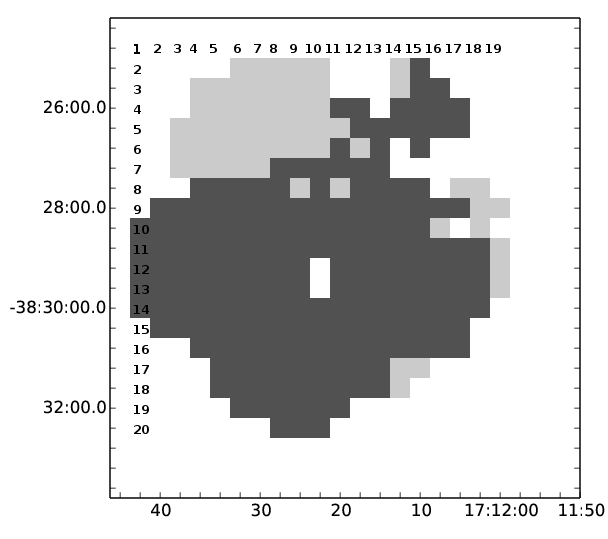} 
\caption{Spatial distribution of velocity profiles with single and double component. Each region represent the same as in Figure \ref{Fig2} with a size of 20$\times$20 pixeles. Profiles with single component are in dark gray and profiles with double velocity components are in light gray.}
\label{Fig3} 
\end{center}
\end{figure}

\begin{figure}[htp]\centering
\begin{center}
\includegraphics[width=1.3\columnwidth, angle =90 ]{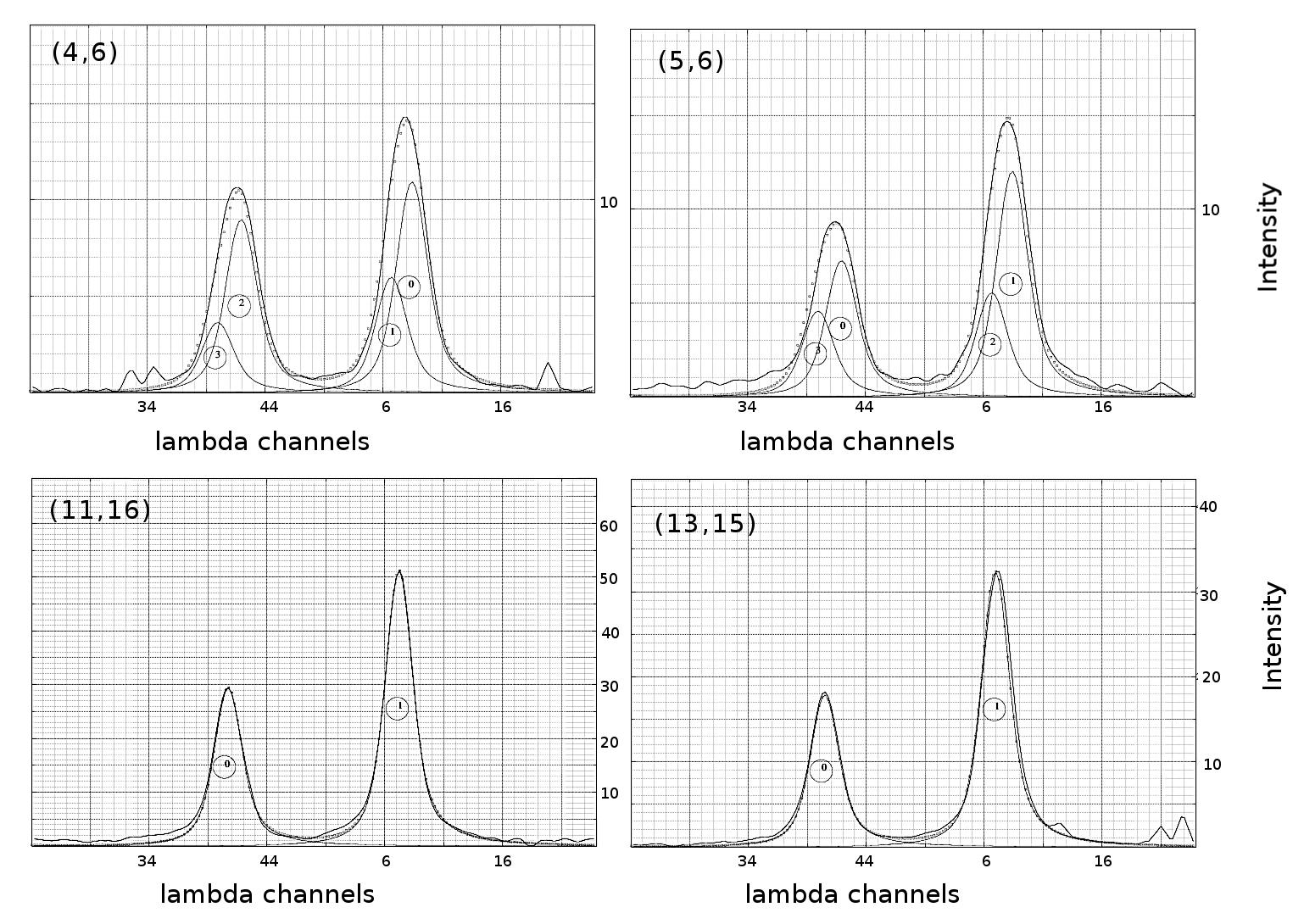} 
\caption{[SII] radial profiles of four regions obtained with Fabry-Perot. In top-left it is presented the (x,y) position of the extracted integrated velocity profile according to Figure \ref{Fig2}. Top profiles are composite, bottom profiles are single. The profiles were integrated  over boxes with a 20$\times$20 pixel size.  The x-axis in the profiles is given in lambda channels and y-axis is the intensity in arbitrary units. Both [SII] lines at 6717 $\rm \AA$ and 6731 $\rm \AA$ are detected. Decomposition of each profile is indicated in thin lines. Resulting profile is shown as hollow circles and with numbers. Dotted line represent the sum of all fitted components.}
\label{Fig4} 
\end{center}
\end{figure}

\begin{figure*}\centering
\includegraphics[width=1\columnwidth]{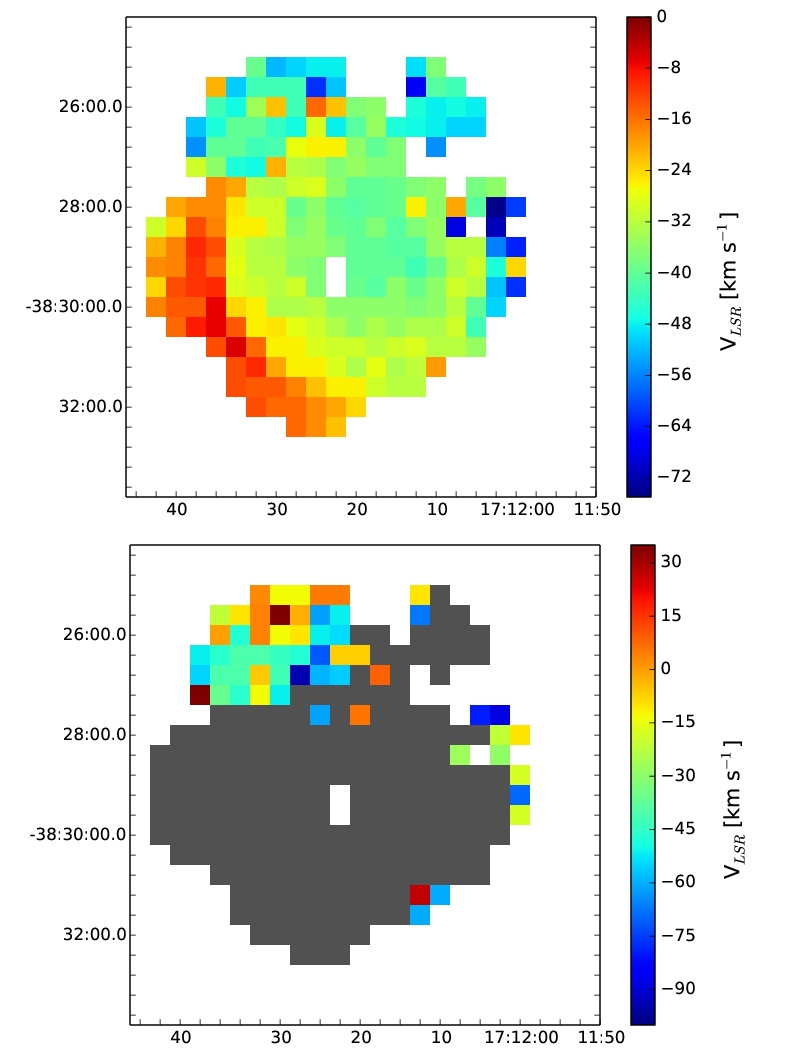} 
\caption{Velocity maps of the components obtained from
the fit profiles. Top panel: velocity map of the $V_{main}$ component. Bottom panel:  velocity map of the V$_{sec}$ component. Color gray  represents the regions with a single velocity component.}
\label{Fig5} 
 \end{figure*}

 \begin{figure}[htp]\centering
\begin{center}
\includegraphics[width=\columnwidth]{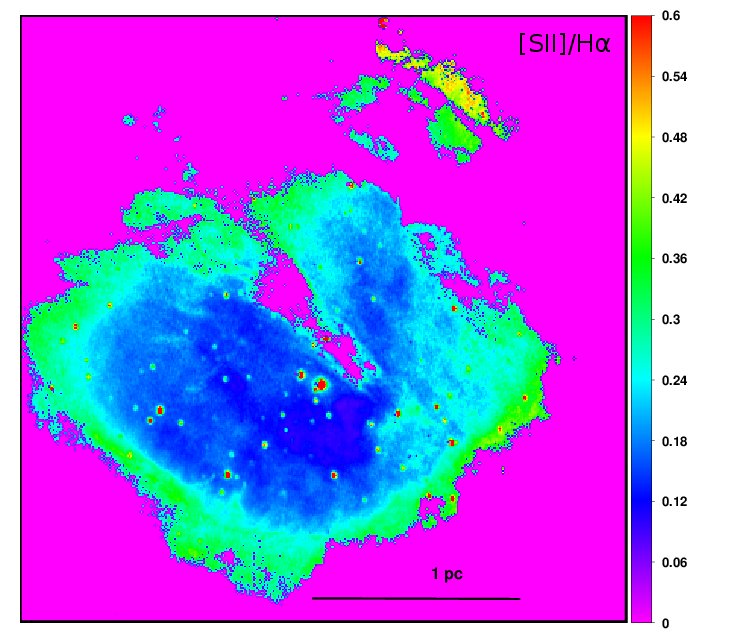} 
\caption{[SII]/H$\alpha$ line-ratio map. The map is oriented such that north is up and east is to the left.}
\label{Fig6} 
\end{center}
\end{figure}

\begin{figure}[htp]\centering
\begin{center}
\includegraphics[width=0.8\columnwidth]{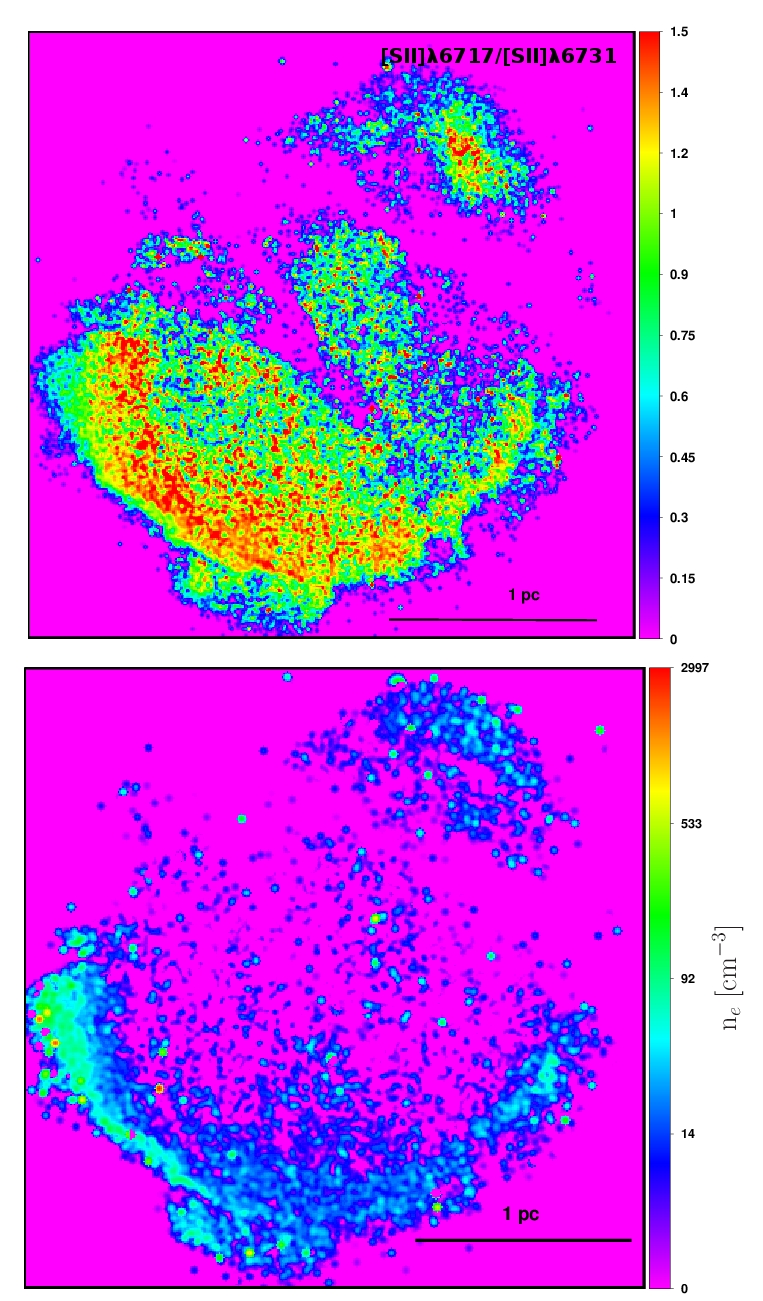}
\caption{\textit{Top panel}: ${\rm I}({\rm [SII]}\, \lambda6717)/{\rm I}({\rm [SII]}\, \lambda6731)$ ratio. \textit{Bottom panel}: Electron density map in
particles per cm$^{-3}$. The maps are oriented such that north is up and east is to the left.}
\label{fig:nelec}
\end{center}
\end{figure}

 \begin{figure*}[t!]
\begin{center}
\includegraphics[width=\columnwidth]{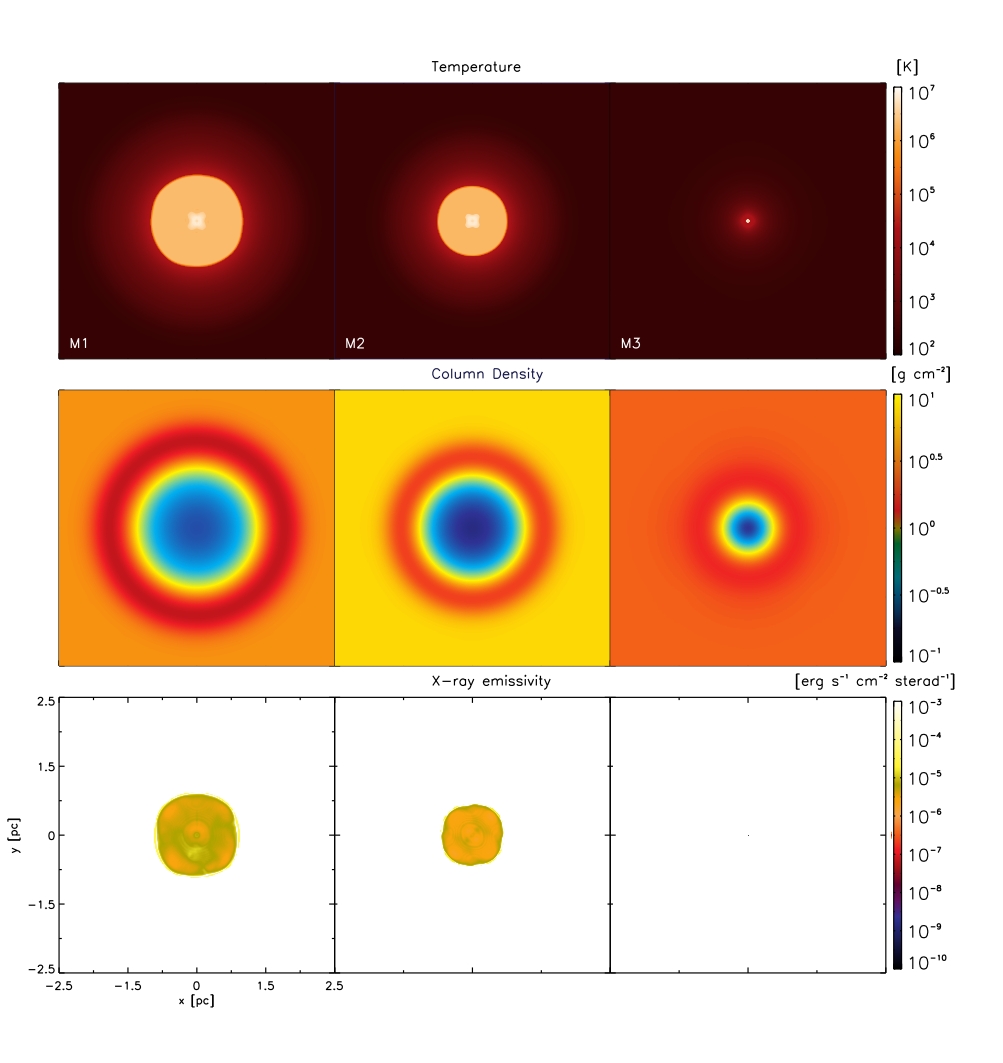}
\caption{Temperature, column density and intrinsic X-ray emission maps, upper, middle and lower panels respectively at evolutionary time of 400 kyr. The results of 
model M1, M2 and M3 are presented.}
\label{fig:nummaps}
\end{center}
\end{figure*}

  \begin{figure*}[t!]
\begin{center}
\includegraphics[width=0.7\columnwidth]{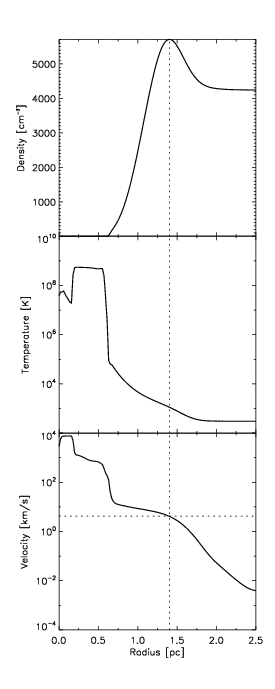}
\caption{Density, temperature, and velocity, upper, middle and lower panels respectively. As we explained in the text, the region of higher values  for the density (marked with a dotted line) correspond to lower values to temperature and velocity and conversely. Moreover, the velocity value obtained in our numerical simulations ($\sim$4 km s$^{-1}$) is in agreement with the observational values reported. }
\label{fig:numgraphs}
\end{center}
\end{figure*}
 
 \begin{table}[!t]\centering
  \setlength{\tabnotewidth}{0.5\columnwidth}
  \tablecols{3}
  \setlength{\tabcolsep}{2.8\tabcolsep}
  \caption{Characteristic of RCW\,120 and its ionizing star} \label{Table1}
 \begin{tabular}{lcl}
    \toprule
\multicolumn{1}{c}{Bubble}& 
\multicolumn{1}{c}{Value}& 
\multicolumn{1}{c}{Reference} \\
\multicolumn{1}{c}{parameters} \\
\midrule
Names							&	Sh2-3	&	\citet{Sharpless1959}	\\
Distance [kpc]					&	1.34 	&	\citet{Zavagno2007}	\\
Radius [pc]						&	1.9 	&	\citet{Anderson2015}	\\
n$_0$ [cm$^{-3}$]				&	2000-6000 	&	\citet{Anderson2015}	\\
n(H$_2$) [cm$^{-2}$]				&	1.4$\times$10$^{22}$ 	&	\citet{Torii2015}	\\
n(H$_2$)$_{\rm ring}$ [cm$^{-2}$]			&	3.22$\times$10$^{22}$ 	&	\citet{Torii2015}	\\
M$_{\rm ring}$ [M$_{\odot}$]			&	3100 	&	\citet{Torii2015}\\
								&	2000 	&	\citet{Deharveng2009}	\\
M(H\,II) [M$_{\odot}$]				&	54 	&	\citet{Zavagno2007}	\\
r$_{\rm ring}$ [pc]					&	1.7 	&	 \citet{Torii2015}	\\
t [Myr]							&	0.4	&	\citet{Torii2015};	\\
								&		&	\citet{Zavagno2007}	\\
\midrule\midrule  			
ionizing star  &		&		\\
parameters &     & \\
\midrule 
Age		[Myr]			&	5 	&	\citet{Martins2010}	\\
B [mag]				&	11.93 	&	\citet{Avedisova-Kondratenko1984}	\\
V [mag]				&	10.79 	&	\citet{Avedisova-Kondratenko1984}	\\
J [mag]				&	8.01 	&	\citet{Avedisova-Kondratenko1984}	\\
H [mag]				&	7.78 	&	\citet{Martins2010}	\\
K [mag]				&	7.52 	&	\citet{Martins2010}\\
M$_*$ [M$_{\odot}$]			&	30 	&	\citet{Martins2010}		\\
log(L$_*$/[L$_{\odot}$])	&	5.07	&	\citet{Martins2010}	\\
t$_{\rm eff}$ [K]							&	37500 &	\citet{Martins2010}	\\
$\dot{\rm M}$ [M$_{\odot}$ yr$^{-1}$]		&	1.55$\times$10$^{-7}$ 	&		\citet{Mackey2015};	\\
						&		 & \citet{Martins2010}	\\ 
L$_0$ [erg s$^{-1}$]\tabnotemark{a}				&	10$^{38}$ 	&		\citet{Martins2010}\\			
\bottomrule
 
 \tabnotetext{a}{L$_0$ is the ionizing photon luminosity}
 \end{tabular}
\end{table}

\begin{table}[!t]\centering
  \setlength{\tabnotewidth}{0.5\columnwidth}
  \tablecols{3}
  \setlength{\tabcolsep}{2.8\tabcolsep}
  \caption{Observational and Instrumental Parameters} \label{Table2}
 \begin{tabular}{lcl}
    \toprule
Parameter & \multicolumn{1}{c}{Value}\\
\midrule
    Telescope								& 2.1m (OAN, SPM) \\
    Instrument								& PUMA \\
    Detector								&  Marconi CCD\\
    Scale plate 							& 0''.33/pix \\
    Binning 								& 4 \\
    Detector size [pixels]					& 2048 $\times$ 2048 \\
    Central Lambda $\rm \AA$ 				&  6720 \\
    Bandwidth $\rm \AA$ 					&  20\\
    Interference Order  $\rm \AA$ 			&  322 at 6717 $\rm \AA$\\
    Free spectral range [km~s$^{-1}$] 		& 929 \\
    Exposure time calibration cube			& 0.5 s/channel \\
    Exposure time object cube 				&  120 s/channel\\
\bottomrule
 \end{tabular}
\end{table}

\begin{table}[!t]\centering
  \setlength{\tabnotewidth}{0.5\columnwidth}
  \tablecols{3}
  \setlength{\tabcolsep}{2.8\tabcolsep}
  \caption{Initial conditions of the numerical simulations} \label{Table3}
 \begin{tabular}{lcccccl}
    \toprule
\multicolumn{1}{c}{ Model } &
\multicolumn{1}{c}{ v$_{\rm term}$} & 
\multicolumn{1}{c}{$\dot{\rm M}$}& 
\multicolumn{1}{c}{n$_0$} &
\multicolumn{1}{c}{T$_0$}&
\multicolumn{1}{c}{Log$_{10}$ L$_w$} \\
 \multicolumn{1}{c}{}&
 \multicolumn{1}{c}{[km~s$^{-1}$] } & 
 \multicolumn{1}{c}{[M$_{\odot}$ yr$^{-1}$]}& 
 \multicolumn{1}{c}{[cm$^{-3}$]}&
 \multicolumn{1}{c}{[K]}\\
 
\midrule
   M1 &  2313 & $2.7\times10^{-7}$ & 2000 & 100 & 5.4 L$_{\odot}$ \\ 
   M2 &  2313 & $2.7\times10^{-7}$ & 3000 & 100 & 5.4 L$_{\odot}$  \\
   M3 &  2313 & $2.7\times10^{-7}$ & 6000 & 100 & 5.4 L$_{\odot}$  \\
 \bottomrule
 \end{tabular}
\end{table}

\begin{table}[!t]\centering
  \setlength{\tabnotewidth}{0.5\columnwidth}
  \tablecols{3}
  \setlength{\tabcolsep}{2.8\tabcolsep}
  \caption{Numerical results} \label{Table5}
 \begin{tabular}{lcccccl}
    \toprule  
\multicolumn{1}{c}{Model}&
\multicolumn{1}{c}{R$_{shell}$ }& 
\multicolumn{1}{c}{$\Delta$ R$_{shell}$}& 
\multicolumn{1}{c}{L$_{X}$}&
\multicolumn{1}{c}{L$_{X,max}$}\\
\multicolumn{1}{c}{}&
\multicolumn{1}{c}{[pc]}& 
\multicolumn{1}{c}{[pc]}& 
\multicolumn{1}{c}{[x10$^{29}$ erg s$^{-1}$]}&
\multicolumn{1}{c}{[x10$^{29}$ erg s$^{-1}$]}\\
\midrule
   M1 &  2.40    & 0.72& 1.32 & 4.84\\ 
   M2 &  2.12    &   0.68  & 1.61 &  4.88\\
   M3 &  1.52   &    0.53  &  2.66&  5.22\\
 \bottomrule
 \end{tabular}
\end{table}

\clearpage 
 \hspace{3cm}
{}
 

\begin{thebibliography}{}
 
\bibitem[Anderson et al.(2015)]{Anderson2015} Anderson, L.~D., 
Deharveng, L., Zavagno, A., et al.\ 2015, \apj, 800, 101 

\bibitem[Avedisova \& Kondratenko(1984)]{Avedisova-Kondratenko1984} Avedisova, V.~S., \& Kondratenko, G.~I.\ 1984, Nauchnye Informatsii, 56, 59 

\bibitem[Campbell(1984)]{Campbell1984} Campbell, B.\ 1984, \apjl, 282, L27 



\bibitem[Castellanos-Ram\'irez et al.(2015)]{Castellanos2015} Castellanos-Ram\'irez A., Rodr\'iguez-Gonz\'alez A., Esquivel, A., Toledo-Roy, J. C., Olivares, J., Vel\'azquez, P. F., 2015, MNRAS, 450, 2799


\bibitem[Chu(2008)]{Chu2008} Chu, Y.-H.\ 2008, Massive Stars as Cosmic Engines, 250, 341 



\bibitem[Chu (2016)]{Chu2016} Chu You-Hua 2016 J. Phys.: Conf. Ser. 728 032007
  
\bibitem[Chu \& Mac Low(1990)]{Chu-MacLow1990} Chu, Y.-H., \& Mac Low, M.-M.\ 1990, \apj, 365, 510 

\bibitem[Chu et al.(1983)]{Chu1983} Chu, Y.-H., Treffers, R.~R., \& Kwitter, K.~B.\ 1983, \apjs, 53, 937 

\bibitem[Chu(1991)]{Chu1991} Chu, Y.~H.\ 1991, Wolf-Rayet Stars and Interrelations with Other Massive Stars in Galaxies, 143, 349 

\bibitem[Chu \& Mac Low(1996)]{Chu-MacLow1996} Chu, Y.-H., \& Mac Low, M.-M.\ 1996, Roentgenstrahlung from the Universe, 241 

\bibitem[Court{\`e}s(1989)]{Courtes1989} Court{\`e}s, G.\ 1989, IAU Colloq.~120: Structure and Dynamics of the Interstellar Medium, 350, 80 

\bibitem[Crowther(2007)]{Crowther2007} Crowther, P.~A.\ 2007, \araa, 45, 177 


\bibitem[Deharveng et al.(2009)]{Deharveng2009} Deharveng, L., Zavagno, A., Schuller, F., et al.\ 2009, \aap, 496, 177 

\bibitem[Dere et al.(1997)]{Dere1997} Dere, K. P., Landi, E., Mason, H. E., Monsignori, Fossi B. C., Young, P. R., 1997, A\&AS, 125, 149

\bibitem[Dyson \& Williams(1980)]{Dyson-Williams1980} Dyson, J.E., \& Williams, D.A. 1980, The Physics of the Interstellar Medium, (New York:John Wiley \& Sons)

\bibitem[Esquivel et al.(2010)]{Esquivel2010}  Esquivel, A., Raga, A. C., Cant\'o, J., Rodr\'iguez-Gonz\'alez, A., L\'opez-C\'amara, D., Vel\'azquez, P. F., De Colle, F., 2010, ApJ, 725, 1466

\bibitem[Friend \& Abbott(1986)]{Friend-Abbott1986} Friend, D.~B., \& Abbott, D.~C.\ 1986, \apj, 311, 701 

\bibitem[Garc{\'{\i}}a-D{\'{\i}}az et al.(2008)]{Garcia2008} Garc{\'{\i}}a-D{\'{\i}}az, M.~T., Henney, W.~J., L{\'{\o}}pez, J.~A., \& Doi, T.\ 2008, \RMAA, 44, 181 

\bibitem[Gvaramadze et al.(2017)]{Gvaramadze2017} Gvaramadze V. V., Mackey J., Kniazev A. Y., Langer N., Chen\'e A.-N., Castro N., Haworth T. J., \& Grebel E. K., 2017, MNRAS, 466, 1857

  \bibitem[Georgelin \& Georgelin(1970)]{Georgelin-Georgelin1970} Georgelin, Y.~P., \& Georgelin, Y.~M.\ 1970, \aaps, 3, 1

\bibitem[Gosset et al.(2005)]{Gosset2005} Gosset, E., Naz{\'e}, Y., Claeskens, J.-F., et al.\ 2005, \aap, 429, 685 

\bibitem[Gruendl et al.(2000)]{Gruendl2000} Gruendl, R.~A., Chu, Y.-H., Dunne, B.~C., \& Points, S.~D.\ 2000, \aj, 120, 2670 

\bibitem[Harper-Clark \& Murray(2009)]{Harper2009} Harper-Clark, E., \& Murray, N.\ 2009, \apj, 693, 1696 


\bibitem[Heckathorn et al.(1982)]{Heckathorn1982} Heckathorn, J.~N., Bruhweiler, F.~C., \& Gull, T.~R.\ 1982, \apj, 252, 230 

\bibitem[Landi et al.(2006)]{Landi2006} Landi, E., Del Zanna, G., Young, P.~R., et al.\ 2006, \apjs, 162, 261 

\bibitem[Le Coarer et al.(1993)]{LeCoarer1993} Le Coarer, E., Rosado, M., Georgelin, Y., Viale, A., \& Goldes, G.\ 1993, \aap, 280, 365 

\bibitem[Mackey et al.(2015)]{Mackey2015} Mackey, J., Gvaramadze, V.~V., Mohamed, S., \& Langer, N.\ 2015, \aap, 573, A10 

  \bibitem[Mackey et al.(2016)]{Mackey2016} Mackey J., Haworth T. J., Gvaramadze V. V., Mohamed S., Langer N., Harries T. J., 2016, A\&A, 586, A114
  
\bibitem[Marston et al.(1994)]{Marston1994} Marston, A.~P., Yocum, D.~R., Garcia-Segura, G., \& Chu, Y.-H.\ 1994, \apjs, 95, 151 

\bibitem[Martins et al.(2010)]{Martins2010} Martins, F., Pomar{\`e}s, M., Deharveng, L., Zavagno, A., \& Bouret, J.~C.\ 2010, \aap, 510, A32 

\bibitem[McCall et al.(1985)]{McCall1985} McCall, M.~L., Rybski, P.~M., \& Shields, G.~A.\ 1985, \apjs, 57, 1 

\bibitem[Miller \& Chu(1993)]{Miller1993} Miller, G.~J., \& Chu, Y.-H.\ 1993, \apjs, 85, 137 

\bibitem[Moore et al.(2000)]{Moore2000} Moore, B.~D., Hester, J.~J., \& Scowen, P.~A.\ 2000, \aj, 119, 2991 

\bibitem[Oey(1996)]{Oey1996} Oey, M.~S.\ 1996, \apj, 465, 231 
 
\bibitem[Prinja et al.(1990)]{Prinja1990} Prinja, R.~K., Barlow, M.~J., \& Howarth, I.~D.\ 1990, \apj, 361, 607 

\bibitem[Raga et al.(2012)]{Raga2012} Raga, A. C., Cant\'o, J., \& Rodr\'iguez, L. F. 2012b, RMxAA, 48, 199
  
\bibitem[Reyes-Iturbide et al.(2014)]{Reyes2014} Reyes-Iturbide, J., Rosado, M., Rodr{\'{\i}}guez-Gonz{\'a}lez, A., et al.\ 2014, \aj, 148, 102 

\bibitem[Rodriguez et al.(1988)]{Rodriguez1988} Rodr\'iiguez, L.~F., Canto, J., \& Moran, J.~M.\ 1988, \apj, 333, 801 

\bibitem[Rodr{\'{\i}}guez-Gonz{\'a}lez et al.(2011)]{Rodriguez2011}  Rodr{\'{\i}}guez-Gonz{\'a}lez, A., Vel{\'a}zquez, P.  ~F., Rosado, M., et al. \ 2011, \apj, 733, 34

\bibitem[Rogers \& Pittard(2014)]{Rogers2014} Rogers, H., \& Pittard, J.~M.\ 2014, \mnras, 441, 964 



\bibitem[Rosado et al.(1995)]{Rosado1995} Rosado, M., Langarica, 
R., Bernal, A., et al.\ 1995, Revista Mexicana de Astronomia y Astrofisica 
Conference Series, 3, 263

\bibitem[Russeil(2003)]{Russeil2003} Russeil, D.\ 2003, \aap, 397, 133 

\bibitem[Sharpless(1959)]{Sharpless1959} Sharpless, S.\ 1959, \apjs, 4, 257 

\bibitem[Spitzer(1978)]{Spitzer1978} Spitzer, L. 1978, Physical Processes in the Intestellar Medium (New York: Wiley
Interscience)

\bibitem[Sternberg et al.(2003)]{Sternberg2003} Sternberg, A., Hoffmann, T. L., \& Pauldrach, A. W. A. 2003, ApJ, 599, 1333

\bibitem[Tenorio-Tagle(1979)]{TenorioTagle1979} Tenorio-Tagle, G., 1979,
  A\&A,71,59.

\bibitem[Tinoco-Arenas et al.(2015)]{Tinoco2015} Tinoco Arenas, A., Gonz\'alez Bol\'ivar, M., Medina Cobarrubias, R., \& Raga, A. C. 2015, RMxAA, 51, 239
  
\bibitem[Toal{\'a} et al.(2016)]{Toala2016} Toal{\'a}, J.~A., Guerrero, M.~A., Chu, Y.-H., et al.\ 2016, \mnras, 456, 4305 

\bibitem[Toledo-Roy et al.(2014)]{Toledo-Roy2014} Toledo-Roy, J. C., Vel\'azquez, P. F., Esquivel, A., Giacani, E., 2014, MNRAS, 437, 898

\bibitem[Torii et al.(2015)]{Torii2015} Torii, K., Hasegawa, K., Hattori, Y., et al.\ 2015, \apj, 806, 7 

\bibitem[Toro et al.(1994)]{Toro1994} Toro, E. F., Spruce, M., \& Speares, W. 1994, Shock Waves, 4, 25
  
\bibitem[Valdez-Guti{\'e}rrez et al.(2001)]{Valdez2001} Valdez-Guti{\'e}rrez, M., Rosado, M., Georgiev, L., Borissova, J., \& Kurtev, R.\ 2001, \aap, 366, 35 

\bibitem[Weaver et al.(1977)]{Weaver1977} Weaver, R., McCray, R., Castor, J., Shapiro, P., \& Moore, R.\ 1977, \apj, 218, 377 

\bibitem[Wrigge et al.(2005)]{Wrigge2005} Wrigge, M., Chu, Y.-H., Magnier, E.~A., \& Wendker, H.~J.\ 2005, \apj, 633, 248 

\bibitem[Zavagno et al.(2007)]{Zavagno2007} Zavagno, A., Pomar{\`e}s, M., Deharveng, L., et al.\ 2007, \aap, 472, 835 

\bibitem[Zavagno et al.(2010)]{Zavagno2010} Zavagno, A., Russeil, D., Motte, F., et al.\ 2010, \aap, 518, L81 

\bibitem[Zhang et al.(2014)]{Zhang2014} Zhang, N. -X., Chu, Y. -H., Williams, R. ~M., et al. \ 2014, \apj, 792, 58

\end{thebibliography}
 \end{document}